\documentclass[conference]{IEEEtran}
\usepackage{balance}
\usepackage[T1]{fontenc}
\usepackage[utf8]{inputenc}
\usepackage{amssymb}
\usepackage[pdftex]{graphicx}
\usepackage{mathtools}
\usepackage{mathabx}
\usepackage{array}
\usepackage{mathtools}
\usepackage[noadjust]{cite}
\usepackage{glossaries}
\usepackage{algorithm}
\usepackage{algpseudocode}
\usepackage{xcolor}
\usepackage{tikz}
\usepackage{pgfplots}
\usepackage{makecell}
\usepgfplotslibrary{fillbetween}
\usepackage{balance}

\makeatletter
\let\MYcaption\@makecaption
\makeatother

\usepackage[font=footnotesize]{subcaption}

\makeatletter
\let\@makecaption\MYcaption
\makeatother
\algnewcommand{\LineComment}[1]{\(\triangleright\) #1}

\newacronym{AWGN}{AWGN}{additive white Gaussian noise}
\newacronym{BLER}{BLER}{block error rate}
\newacronym{CE}{CE}{cross-entropy}
\newacronym{DL}{DL}{deep learning}
\newacronym{iid}{i.i.d.\@}{independent and identically distributed}
\newacronym{ML}{ML}{maximum likelihood}
\newacronym{MIMO}{MIMO}{multiple-input multiple-output}
\newacronym{NN}{NN}{neural network}
\newacronym{RBF}{RBF}{Rayleigh block-fading}
\newacronym{ReLU}{ReLU}{rectified linear unit}
\newacronym{SNR}{SNR}{signal-to-noise ratio}
\newacronym{SGD}{SGD}{stochastic gradient descent}
\newacronym{wrt}{w.r.t.\@}{with respect to}
\newacronym{FPGA}{FPGA}{field-programmable gate array}
\newacronym{GPU}{GPU}{graphics processing unit}
\newacronym{ASIC}{ASIC}{application-specific integrated circuit}
\newacronym{NLP}{NLP}{natural language processing}
\newacronym{LC}{LC}{learning-compression}
\newacronym{DC}{DC}{direct compression}
\newacronym{OFDM}{OFDM}{orthogonal frequency-division multiplexing}
\newacronym{NC}{NC}{not compressed}
\renewcommand{\vec}[1]{\mathbf{#1}}
\newcommand{\vecs}[1]{\boldsymbol{#1}}

\newcommand{\pv}{\vec{p}}

\newcommand{\xv}{\vec{x}}
\newcommand{\yv}{\vec{y}}

\newcommand{\zerov}{\vec{0}}

\newcommand{\lambdav}{\vecs{\lambda}}

\newcommand{\thetav}{\vecs{\theta}}
\newcommand{\psiv}{\vecs{\psi}}

\newcommand{\Bc}{{\cal B}}
\newcommand{\Cc}{{\cal C}}

\newcommand{\CC}{\mathbb{C}}
\newcommand{\MM}{\mathbb{M}}

\newcommand{\RR}{\mathbb{R}}
\newcommand{\ZZ}{\mathbb{Z}}

\newcommand{\LB}{\left(}
\newcommand{\RB}{\right)}
\newcommand{\LP}{\left\{}
\newcommand{\RP}{\right\}}

\newcommand{\EE}{{\mathbb{E}}}

\newcommand\norm[1]{\left\lVert#1\right\rVert}
\newcommand\abs[1]{\mathopen|#1\mathclose|}

\begin{document}
\title{Towards Hardware Implementation of Neural Network-based Communication Algorithms}
\author{
\IEEEauthorblockN{Fayçal Ait Aoudia and Jakob Hoydis}
\IEEEauthorblockA{Nokia Bell Labs\\\{faycal.ait\_aoudia, jakob.hoydis\}@nokia-bell-labs.com
}}
\maketitle

\begin{abstract}
There is a recent interest in \gls{NN}-based communication algorithms which have shown to achieve (beyond) state-of-the-art performance for a variety of problems or lead to reduced implementation complexity.
However, most work on this topic is simulation based and implementation on specialized hardware for fast inference, such as \glspl{FPGA}, is widely ignored.
In particular for practical uses, \gls{NN} weights should be quantized and inference carried out by a fixed-point instead of floating-point system, widely used in consumer class computers and \glspl{GPU}.
Moving to such representations enables higher inference rates and complexity reductions, at the cost of precision loss.
We demonstrate that it is possible to implement \gls{NN}-based algorithms in fixed-point arithmetic with quantized weights at negligible performance loss and with hardware complexity compatible with practical systems, such as \glspl{FPGA} and \glspl{ASIC}.
\end{abstract}
\glsresetall

\section{Introduction}

 \begin{figure}[b]
    \centering
      \includegraphics[width=\linewidth]{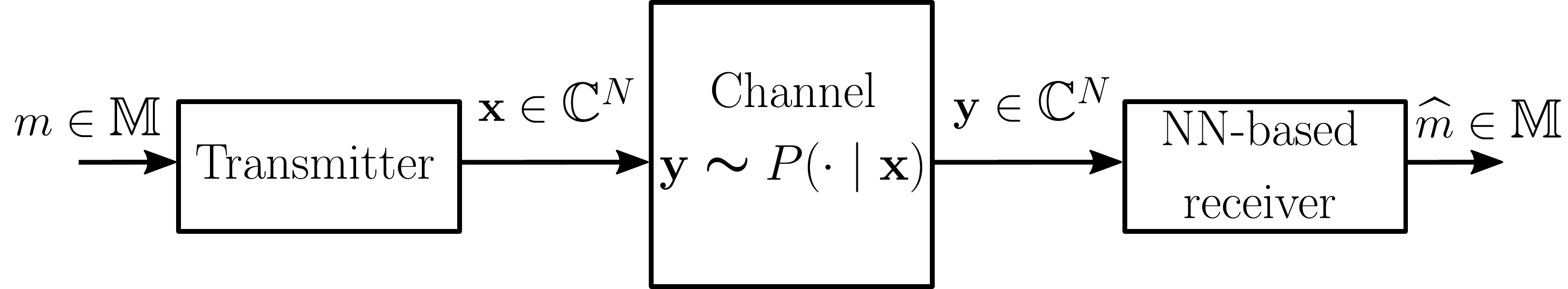}
    \caption{A communication system leveraging an \gls{NN}-based receiver}
\label{fig:big_pic}
\end{figure}

Inspired by the success of \gls{DL} in various fields such as computer vision and \gls{NLP}, \gls{NN}-based communication systems have gained a lot of attention recently.
Approaches leveraging \gls{DL} have lately emerged for channel coding~\cite{8259241}, \gls{MIMO} systems~\cite{8227772}, \gls{OFDM}~\cite{8052521}, and so forth.
However, most of these contributions are simulation-based, and only a few have considered hardware implementation of \gls{DL}-based approaches as well as the issues it raises.
Indeed, one of the biggest obstacle to the use of \glspl{NN} is their high memory requirement and computational complexity.
Hardware acceleration is needed to achieve reasonable inference time, and most of previous contributions leverage \glspl{GPU}, which come at high monetary and energy cost not viable for communication systems.
Indeed, inference speed on the physical layer are on the order of micro to nanoseconds, which is at least one order of magnitude faster than in other applications, e.g., autonomous cars.

\begin{figure}
	\begin{tikzpicture}
		\begin{axis}[
			ymode=log,
			grid=both,
			grid style={line width=.1pt, draw=gray!10},
			major grid style={line width=.2pt,draw=gray!50},
			minor tick num=5,
			xlabel={SNR (dB)},
			ylabel={BLER},
			legend style={at={(0.1, 0.1)},anchor=south west}
		]
			\addplot[orange, mark=diamond] table [x=snr, y=dc_fxp, col sep=comma] {figs/results_intro.csv};
			\addplot[black!30!green, mark=square] table [x=snr, y=nq, col sep=comma] {figs/results_intro.csv};
			\addplot[blue, mark=triangle] table [x=snr, y=ml, col sep=comma] {figs/results_intro.csv};

			\addlegendentry{Naive compression}
			\addlegendentry{Uncompressed}
			\addlegendentry{Maximum likelihood}

			\end{axis}
		\end{tikzpicture}
    \caption{The uncompressed \gls{NN}-based receiver achieves BLERs close to the one of maximum likelihood detection. However, naive compression of the receiver leads to significant degradation of the BLER.}
    \label{fig:intro_res}
    \vspace{-20pt}
\end{figure}
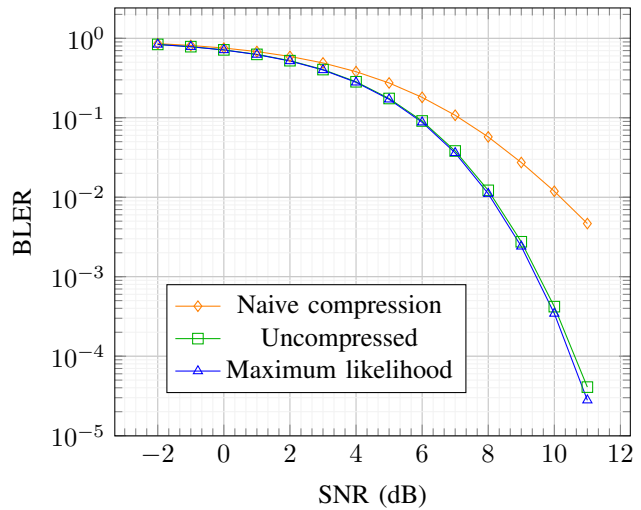

Recently, multiple methods were proposed to compress \glspl{NN} to reduce their complexity, such as weight quantization~\cite{NIPS2017_7163}, weight pruning~\cite{gordon2018morphnet}, or more efficient implementations of the conventional floating-point operators~\cite{johnson2018rethinking}.
Nevertheless, such methods were mostly considered for flagship machine learning tasks, such as image classification and speech recognition.
In this paper, we consider the efficient implementation of \gls{NN}-based communication algorithms in fixed-point arithmetic.
Our aim is not to reduce the memory footprint of the \gls{NN} by learning a compressed representation of the architecture, but its computational complexity as we assume the \gls{NN} is implemented in hardware.
Fixed-point compute units are faster and consume less hardware resources and energy than conventional floating-point units~\cite{NIPS2017_7163,johnson2018rethinking}.
We consider the implementation of an \gls{NN}-based receiver as shown in Fig.~\ref{fig:big_pic}.
However, the approach used in this paper can be applied to a wide variety of \gls{NN}-based communication algorithm, e.g., fully learned transceiver implementations as done in~\cite{8054694}.
The \gls{NN}-based receiver trained with no constraints and uncompressed enables \glspl{BLER} close to the ones of \gls{ML} detection as shown in Fig.~\ref{fig:intro_res}.
In this figure, we assume a transmitter implementing the Agrell~\cite{Agrell16} scheme with eight bit blocks and four channel uses transmitting over an \gls{AWGN} channel.
We aim to quantize the weights of the \gls{NN}-based receiver so that they take values in a finite codebook, enabling a more efficient implementation.
A straightforward approach to quantize the \gls{NN}-based receiver is to first train it with no constraints, and then to quantize its weights.
This naive approach usually leads to significant performance loss as shown in Fig.~\ref{fig:intro_res}, which motivates the use of quantization-aware algorithms.
In this paper, we leverage a two-stages approach to achieve efficient implementation without significant performance loss:
\begin{enumerate}
	\item First, the weights are quantized using the \gls{LC} algorithm~\cite{carreira2017modelquant} so that they take values in a predefined finite codebook.
Training using \gls{LC} is done on regular high precision floating-point arithmetic.
	\item Next, the quantized \gls{NN} is implemented on a fixed-point arithmetic system, less precise than the floating-point system used to train it, but with a reduced complexity.
\end{enumerate}
We show that the quantized \gls{NN}-receiver trained on a 32~bit floating-point arithmetic system but implemented on a 14~bit fixed-point arithmetic system enables \glspl{BLER} close to the ones of \gls{ML} detection, while being more than $60\%$ less complex.

To the best of our knowledge, only a few papers have considered implementations of \gls{NN}-based communication algorithms.
\gls{NN}-based transceivers~\cite{8054694} were implemented on \gls{FPGA} in~\cite{kim2018building}.
The implementation of a \gls{DL}-based modulation classifier on \gls{FPGA} was described in~\cite{electronics7070122}.
However, in neither of these contributions did the authors attempt to reduce the model complexity to make inference more efficient.
In~\cite{teng2018low}, a recurrent \gls{NN} was considered to decode polar codes.
After each training epoch, the weights were quantized in a two-step process: the weights were first rounded to the nearest fixed-point value that can be represented by a predefined number of bits, before being assigned values from a codebook based on the frequency with which rounded weights appear.
The authors showed that weight quantization enables significant decrease of the memory footprint and computational complexity, without significant performance loss.
However, evaluation of the \gls{NN} in fixed-point arithmetic was not performed.

\textbf{Notations:}
Boldface upper- and lower-case letters denote matrices and column vectors, respectively.
$\RR$ and $\CC$ respectively denote the sets of real and complex numbers.
$\ZZ$ denotes the set of integer.

\section{Background on \gls{NN} compression}

\subsection{Fixed-point arithmetic}

Conventional hardware used in machine learning, such as \glspl{GPU}, rely on floating-point arithmetic.
With this scheme, real numbers that can be represented exactly are of the form
\begin{equation}
	z = \operatorname{sign}(z) s 2^{e}
\end{equation}
where $s$ and $e$ are integers called the significand and the exponent, respectively, and $\operatorname{sign}$ is the sign function.
The numbers of bits used for the significand and the exponent control the range and precision of the representation.
A floating-point number is stored as the sign bit, the exponent field, and the significand field.
A key feature of floating-point representation is that it does not form a uniformly-spaced grid, as the spacing between consecutive numbers grows with the exponent.
This enables the representation of numbers of widely different orders of magnitude.
The main drawback of floating-point arithmetic is the complexity of its compute units, which require a lot of hardware resources, energy, and time compared to other schemes, such as fixed-point arithmetic.
Indeed, performing an operation (such as an addition or multiplication) usually requires preprocessing of the operands and post-processing of the result if their exponents are different.

\begin{figure}
	\centering
    \includegraphics[width=0.85\linewidth]{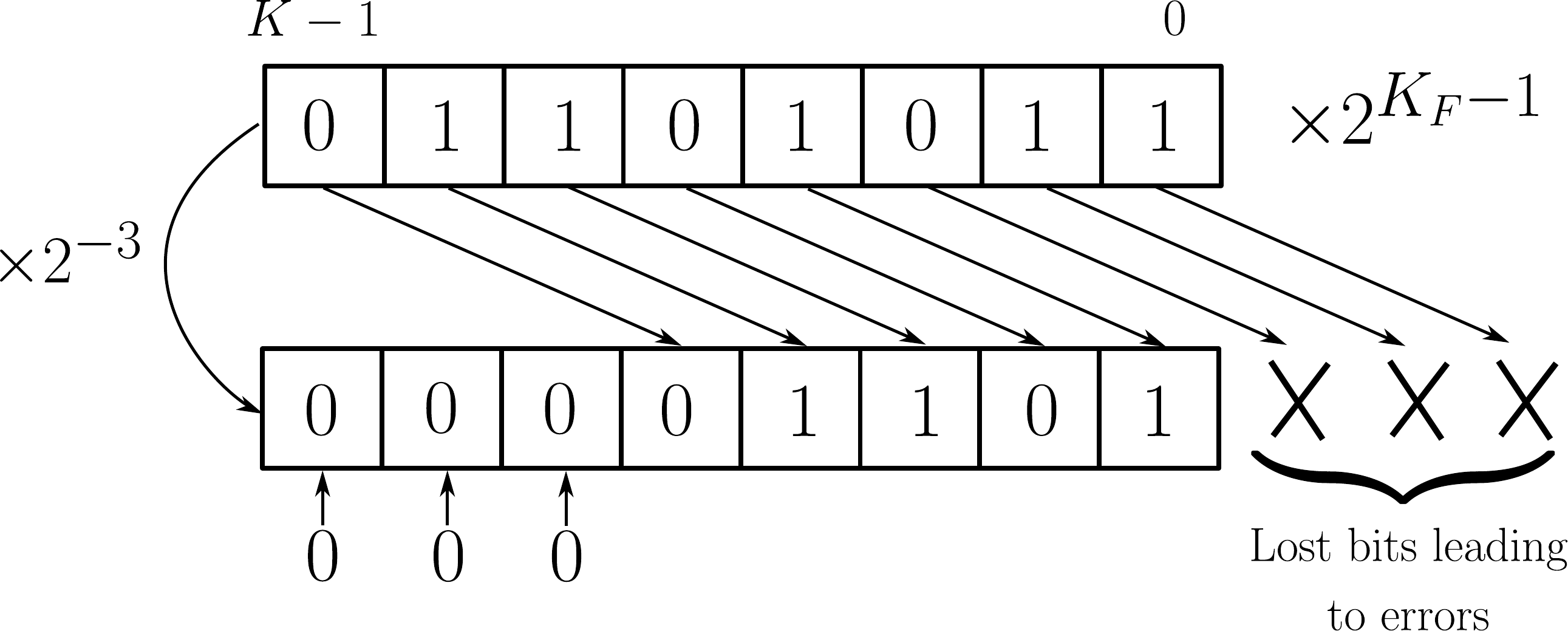}
    \caption{Multiplication by powers of two can be implemented as simple bit-shifts in fixed-point arithmetic.}
	\label{fig:bit_shift}
	\vspace{-10pt}
\end{figure}


Regarding fixed-point arithmetic, only real numbers that can be written as
\begin{equation} \label{eq:fxp}
	z = \operatorname{sign}(z) \LB \sum_{i=0}^{K_I-1} \Bc(z)_i 2^i + \sum_{i=1}^{K_F} \Bc(z)_{-i} 2^{-i} \RB
\end{equation}
can be represented.
$K_I$ and $K_F$ are non-negative integers that correspond to the number of bits of the integer and fractional parts, respectively.
Notice that (\ref{eq:fxp}) corresponds to writing $z$ in the binary numeral system and constraining the number of bits allowed for the integer and fractional parts to finite values.
One can see that, contrary to floating-point representation, representable numbers form a uniformly-spaced grid whose range and precision are controlled by $K_I$ and $K_F$, respectively.
Two consecutive numbers are spaced by $2^{-K_F}$.
A fixed-point number is typically stored using $K = K_I + K_F + 1$ bit (an additional bit is required to handle negative numbers), as shown in Fig.~\ref{fig:bit_shift}.
The number is represented as an $K$~bit integer, with an implicit factor $2^{-K_F}$ that does not need to be stored as it is fixed.
With regard to complexity, fixed-point operators are typically of low complexity compared to floating-point operators as no additional processing steps needs to be taken, which motivates their use to reduce the memory footprint and computational requirements of \glspl{NN}, e.g.~\cite{gupta2015deep}.
We adopt the same approach in this paper, by implementing an \gls{NN}-based receiver in fixed-point arithmetic.

\subsection{The LC algorithm} \label{sec:lc}

\begin{algorithm}
\caption{\gls{LC} algorithm.}
\label{alg:lc}
\begin{algorithmic}[1]
\State $\psiv \gets \underset{\psiv}{\arg\min}\LB L(\psiv) \RB$ \label{lst:init_psi}
\State $\widehat{\psiv} \gets \underset{\widehat{\psiv} \in \Cc^P}{\arg\min}\LB \norm{\widehat{\psiv} - \psiv}^2 \RB$ \label{lst:init_qpsi}
\State $\lambdav \gets \zerov$
\For {$\mu = \mu^{(0)}, \mu^{(1)},\dots$}
	\State Set $\psiv$ by solving (\ref{eq:lc_lrn}) \label{lst:learn}
	
	\State Set $\widehat{\psiv}$ by solving (\ref{eq:lc_cmp})

	\State $\lambdav \gets \lambdav - \mu\LB \psiv - \widehat{\psiv} \RB$

	\State \textbf{Stop if} $\norm{\psiv - \widehat{\psiv}}$ is small enough
\EndFor
\end{algorithmic}
\end{algorithm}

Moving to fixed-point arithmetic is not the only way to reduce the resources required by an \gls{NN}.
Another approach used in this paper conjointly with implementation in fixed-point arithmetic is quantization of the weights.
Quantizing the weights of an \gls{NN} means forcing them to take values in a discrete codebook.

It is well-known that multiplication is significantly more computationally demanding than addition in fixed-point arithmetic.
Indeed, a fixed-point multiplication of $K$~bit operands requires up to $K$~bit shifts and $K-1$ additions.
By forcing the weights to take values in a well-chosen codebooks, the cost of multiplications can be drastically reduced.
For example, using the codebook $\{-1, 0, 1\}$ reduces multiplications to zeroing or sign changes.
Also, choosing the codebook to be a set of powers of two reduces multiplication to bit shifts in fixed-point arithmetic, as multiplication by $2^q$ is equivalent to moving the radix point $q$ digits to the left or right depending on the sign of $q$, as illustrated in Fig.~\ref{fig:bit_shift}.

A key question is how to train an \gls{NN} while forcing its weights to take values in a given codebook.
Let us denote by $f_{\psiv}$ the mapping implemented by an \gls{NN} with parameters $\psiv \in \RR^P$, $L(\psiv)$ the loss function, $\Cc$ the quantization codebook, and $\widehat{\psiv} \in \Cc^P$ the quantized weights.
A naive approach is to first train the \gls{NN} with no constraints on its weights by solving $\arg\min_{\psiv} \LB L(\psiv) \RB$ and then to quantize the model by choosing for each weight its closest value in the codebook by solving $\arg\min_{\widehat{\psiv} \in \Cc^P} \LB \norm{\widehat{\psiv} - \psiv}^2_2 \RB$.
This simple approach, referred to as \gls{DC}, typically does not lead to satisfactory results.
To circumvent this issue, compression-aware algorithms are typically used.
One such algorithm is \gls{LC}~\cite{carreira2017modelquant}, in which compression of an \gls{NN} is considered as a constrained optimization problem, solved by applying alternating optimization to the augmented Lagrangian.
\gls{LC} is guaranteed to converge to a local optimum in some cases~\cite[Section~3.2]{carreira2017modelquant}.
Each iteration of \gls{LC} performs two steps, a learning step and a quantization step.
The learning step updates the unquantized weights $\psiv$ by solving
\begin{equation} \label{eq:lc_lrn}
	\underset{\psiv}{\arg\min} \LB L(\psiv) + \frac{\mu}{2} \norm{\psiv - \widehat{\psiv} - \frac{1}{\mu}\lambdav}^2_2 \RB
\end{equation}
where $\mu$ is a parameter of the algorithm, which is increased at each iteration following a predefined schedule, and $\lambdav$ is the Lagrange multiplier estimate.
One can see that a regularization term is added to the loss, which ensures that the unquantized weights stay close to $\widehat{\psiv} + \frac{1}{\mu}\lambdav$.
The compression step updates the quantized weights $\widehat{\psiv}$ by solving
\begin{equation} \label{eq:lc_cmp}
	\underset{\widehat{\psiv} \in \Cc^P}{\arg\min} \LB \norm{\psiv - \frac{1}{\mu}\lambdav - \widehat{\psiv}}^2_2 \RB
\end{equation}
which corresponds to the quantization of $\psiv - \frac{1}{\mu}\lambdav$ to the codebook $\Cc$.
The \gls{LC} algorithm is depicted in Algorithm~\ref{alg:lc}.
The parameter $\mu$ is typically increased following a multiplicative schedule $\mu^{(k)} = a\mu^{(k-1)}$, where $\mu^{(k)}$ is the value of $\mu$ at the $k$th iteration and $a$ is a parameter of the algorithm larger than one.
$\psiv$ and $\widehat{\psiv}$ are initialized by training the \gls{NN} without any constraints and then quantizing the weights (lines~\ref{lst:init_psi} and~\ref{lst:init_qpsi}).
In practice, the learning step (line~\ref{lst:learn}) is approximatively solved using \gls{SGD} or a variant.

\section{Quantization of \gls{NN}-based receiver}

In this section, efficient implementation of an \gls{NN}-based receiver is achieved using a two-stages approach.
First, quantization of the \gls{NN}-based receiver is performed by training it with \gls{LC} on a usual floating-point arithmetic system.
Next, the quantized \gls{NN} is implemented and evaluated on a fixed-point system.
While this paper focuses on an \gls{NN}-based receiver, this approach can be applied to a wide variety of \gls{NN}-based communication algorithms.

\subsection{NN-based receiver}

In a point-to-point communication system, two nodes aim to reliably exchange information over a stochastic channel as shown in Fig.~\ref{fig:big_pic}.
The output of the channel $\yv$ follows a probability distribution conditional to its input $\xv$, i.e., $\yv \sim P(\yv|\xv)$.
The transmitter aims to communicate messages $m$ drawn from a finite set $\MM = \{1,\dots,M\}$, while the task of the receiver is to detect the sent messages $m$ from the received signal $\yv$.
The receiver is implemented as an \gls{NN} (see Fig.~\ref{fig:big_pic}) $f_{\thetav_R}^{(R)} : \CC^N \mapsto \LP \pv \in \RR_+^M | \sum_{i=1}^M p_i = 1 \RP$, where $\thetav_R$ is the set of parameters and $N$ the number of channel uses.
Its purpose is to estimate the conditional probability $P(m|\yv)$, which corresponds to a supervised learning task.
Once trained, the receiver can be deployed for practical use.

A communication system operating over an \gls{AWGN} channel is considered in this work, with $M = 256$ and $N = 4$.
The transmitter implements the Agrell scheme~\cite{Agrell16}, a subset of the E8 lattice designed by numerical optimization to approximately solve the sphere packing problem for $M = 256$ in eight dimensions (corresponding to four channel uses).
Normalization is performed to ensure that $\EE \LP \frac{1}{N} \norm{\xv}^2 \RB = e_s$, where $e_s$ is the energy per complex symbol.

\begin{figure}
	\centering
    \includegraphics[width=0.7\linewidth]{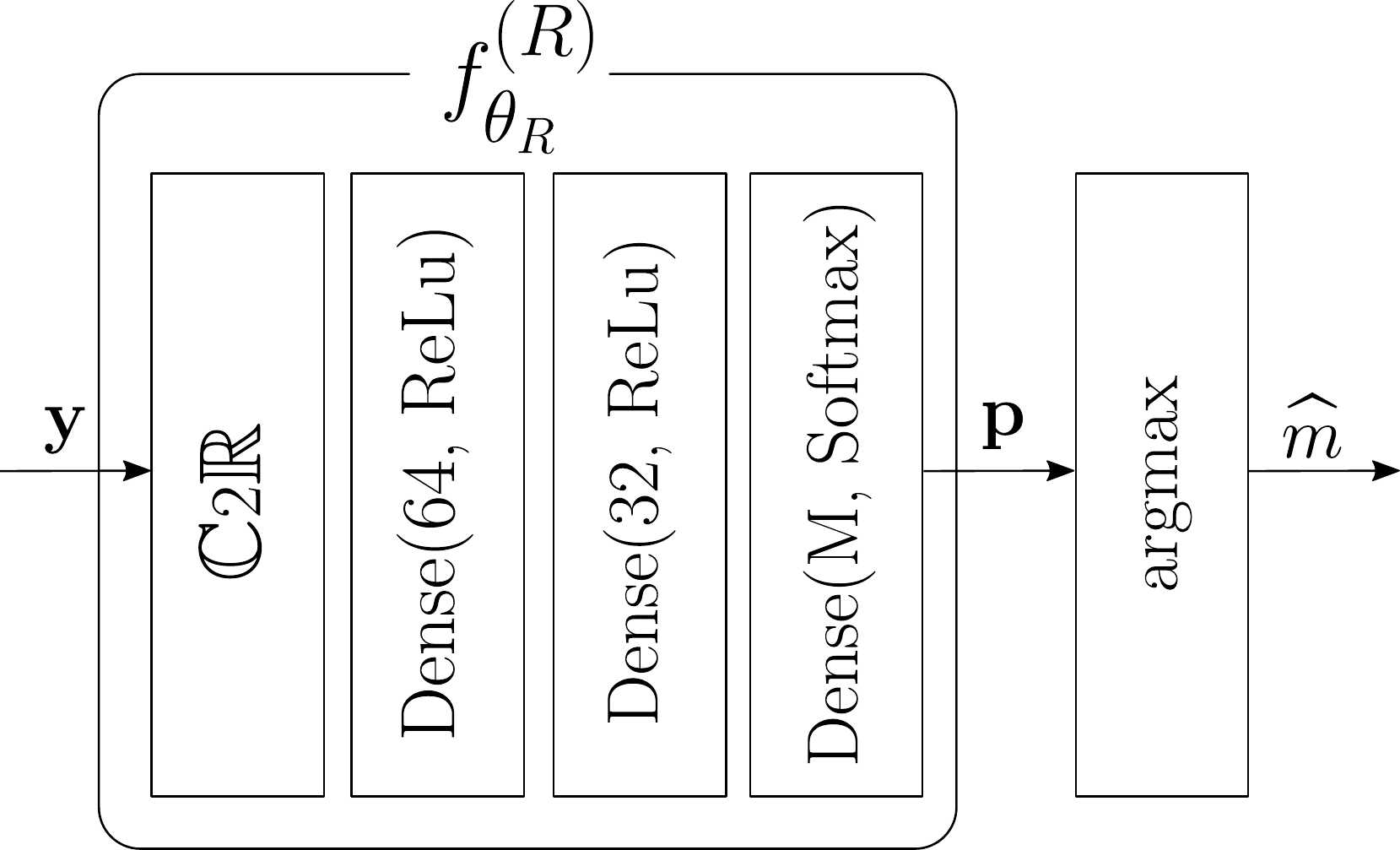}
    \caption{Architecture of the receiver}
	\label{fig:rx_archi}
	\vspace{-10pt}
\end{figure}

\subsection{Architecture of the \gls{NN}-based receiver}

The receiver is implemented by a $\CC2\RR$ layer, mapping the $N$ received complex symbols to $2N$ real numbers, followed by a dense layer of 64 units with ReLu activation, a dense layer of 32 units with ReLu activation, and finally a dense layer of $M$ units with softmax activation, as shown in Fig.~\ref{fig:rx_archi}.
All the dense layers but the last use biases.
Hard decoding is performed by taking the message with highest probability.

Regarding implementation considerations, the ReLu activation was chosen as it requires minimal overhead.
Indeed, its implementation requires neither approximation using a look-up table nor arithmetic operations.
Therefore, it does not incur computational overhead nor arithmetic errors due to approximation.
Moreover, implementation of the output layer softmax activation is not required at deployment, as hard decoding can be performed based on the pre-activations.

\subsection{Weight quantization}

\begin{figure}
	\begin{tikzpicture}
		\begin{axis}[
			ymode=log,
			grid=both,
			grid style={line width=.1pt, draw=gray!10},
			major grid style={line width=.2pt,draw=gray!50},
			minor tick num=5,
			xlabel={SNR (dB)},
			ylabel={BLER},
			legend style={at={(0.1, 0.1)},anchor=south west}
		]
			\addplot[orange, mark=diamond] table [x=snr, y=dc, col sep=comma] {figs/results.csv};
			\addplot[black, mark=o] table [x=snr, y=lc, col sep=comma] {figs/results.csv};
			\addplot[black!30!green, mark=square] table [x=snr, y=nq, col sep=comma] {figs/results.csv};
			\addplot[blue, mark=triangle] table [x=snr, y=ml, col sep=comma] {figs/results.csv};

			\addlegendentry{DC -- $K_F = 8\:$bit}
			\addlegendentry{LC -- $K_F = 8\:$bit}
			\addlegendentry{Not quantized}
			\addlegendentry{ML}

			\end{axis}
		\end{tikzpicture}
    \caption{\gls{BLER} achieved by the evaluated schemes}
    \label{fig:results}
    \vspace{-10pt}
\end{figure}
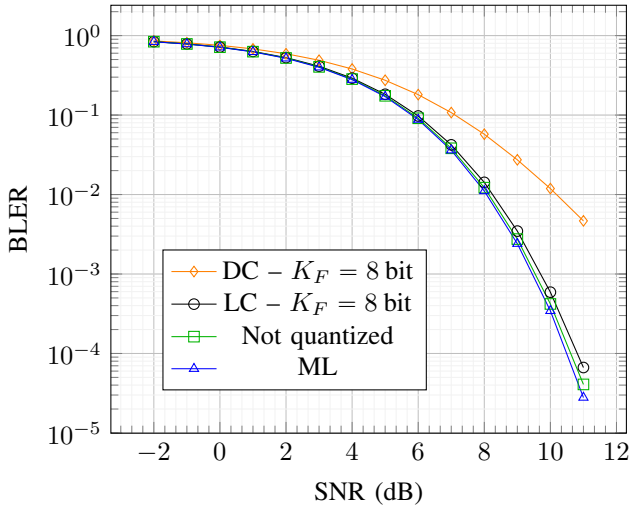

Quantization of the \gls{NN}-based receiver was done by training the \gls{NN} using the \gls{LC} algorithm presented in Section~\ref{sec:lc}, on usual \glspl{GPU} with floating-point arithmetic.
Different codebooks were used for the weights and biases.
Regarding the weights, the codebook was
\begin{equation}
	\Cc_W = \{0, \pm2^q~|~q \in \ZZ, \abs{q} < K-1\},
\end{equation}
where $K = K_I + K_F + 1$.
The choice of this codebook was motivated by the much higher complexity of multiplications compared to additions.
Accordingly, with this codebook, all multiplications are reduced to either zeroing or bit shifting.
Moreover, multiplications by $2^q$ with $\abs{q} \geq K-1$ lead to zeroing on a $K~$bit fixed-point system.
Therefore, the codebook was restricted to powers of two with exponent less than $K-1$ in absolute value.
Biases were constraint to take values in the codebook defined by the set of fixed-point numbers with $K_I$~bit for the integer part and $K_F$~bit for the fractional, i.e., the set of real numbers that can be represented as in~(\ref{eq:fxp}).

To evaluate the impact of receiver quantization on the \gls{BLER}, comparison was done between the \gls{ML} receiver, the unquantized, \gls{LC}- and \gls{DC}-quantized \gls{NN}-based receiver.
When quantization was performed, $K_I$ was set to 5 as it was experimentally found to be the smallest value large enough to avoid overflows on a fixed-point system.
Trainings and evaluations were performed for values of $K_F$ of 2, 4, 8, and 12~bit.
The \gls{SNR} is defined as
\begin{equation}
	\text{SNR} \coloneq \frac{\EE\LP\frac{1}{N}\norm{\xv}^2_2\RP}{\sigma^2} = \frac{e_s}{\sigma^2}
\end{equation}
where $\sigma^2$ is the per-complex noise symbol variance, and the equality results from the energy constraint ensured by the transmitter normalization layer.
$\sigma^2$ was set to $-80\:$dB, and the \gls{SNR} was controlled by setting $e_s$.
Evaluations were done using the Tensorflow~\cite{tensorflow2015-whitepaper} framework and training with the Adam~\cite{Kingma15} variant of \gls{SGD}.
Fig.~\ref{fig:results} shows the \gls{BLER} achieved by the compared schemes for \gls{SNR} values ranging from $-2\:$dB to $11\:$dB.
Evaluations were performed on a floating-point system.
Only results with $K_F=8\:$bit are shown for readability, as other values of $K_F$ lead to almost identical \glspl{BLER}.
One can see that quantization using the naive \gls{DC} approach leads to higher error rates than quantization with \gls{LC}.
Moreover, quantizing the \gls{NN}-based receiver using \gls{LC} leads to \glspl{BLER} close to the ones achieved by the not quantized \gls{NN}-based receiver and \gls{ML} detection.

\subsection{Impact of fixed-point arithmetic}

\begin{figure}
	\begin{tikzpicture}
		\begin{axis}[
			ymode=log,
			grid=both,
			grid style={line width=.1pt, draw=gray!10},
			major grid style={line width=.2pt,draw=gray!50},
			minor tick num=5,
			xlabel={SNR (dB)},
			ylabel={BLER},
			legend style={at={(0.1, 0.1)},anchor=south west}
		]
			\addplot[black, mark=o] table [x=snr, y=fxp2, col sep=comma] {figs/results_fxp.csv};
			\addplot[orange, mark=diamond] table [x=snr, y=fxp4, col sep=comma] {figs/results_fxp.csv};
			\addplot[black!30!green, mark=square] table [x=snr, y=fxp8, col sep=comma] {figs/results_fxp.csv};
			\addplot[blue, mark=triangle] table [x=snr, y=flp, col sep=comma] {figs/results_fxp.csv};

			\addlegendentry{Fixed-point - 2 bits}
			\addlegendentry{Fixed-point - 4 bits}
			\addlegendentry{Fixed-point - 8 bits}
			\addlegendentry{Floating-point}

			\end{axis}
		\end{tikzpicture}
    \caption{Impact of erroneous fixed-point arithmetic on \gls{BLER}}
    \label{fig:results_fxp}
    \vspace{-10pt}
\end{figure}
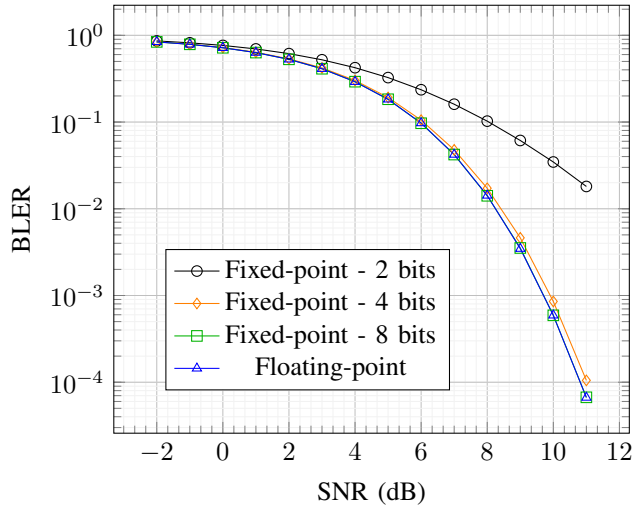

This section investigates the impact on the \gls{BLER} of implementing the quantized \gls{NN}-receiver on a fixed-point arithmetic system.
The quantized \glspl{NN} trained on \glspl{GPU} with \gls{LC} for $K_I = 5$~bit and different values of $K_F$ were implemented on a fixed-point system with the corresponding number of bits allocated to the integer and fractional part.
Fixed-point arithmetic was simulated in Python.
As all the weights are powers of two, all multiplications were reduced to bit shifts.
Implementation of the ReLu activation function is straightforward.
The softmax activation of the output layer was not implemented, as hard decoding can be performed based on the pre-activations.
Fig.~\ref{fig:results_fxp} shows the \gls{BLER} achieved by the receiver for different values of $K_F$.
It can be seen that using only 2~or 4~bit for the fractional part leads to significant increase of the error rate, while using 8~bit (or more) leads to no \gls{BLER} degradation.
These result shows that it is possible to implement the \gls{NN}-based receiver in 14~bits fixed-point arithmetic with no \gls{BLER} degradation, despite the fact that it was trained on a 32~bits floating point arithmetic system.

\subsection{Complexity evaluation}

It was shown in the previous section that an \gls{NN}-based receiver with weights taking as values powers of two and implemented on a fixed-point arithmetic system can achieve \gls{BLER} close to the ones of \gls{ML} detection.
In this section, we compare the computational complexity of the previously evaluated \gls{ML} receiver and quantized \gls{NN}-based receiver.

Regarding the \gls{NN}-based receiver, only multiplications and additions are required.
Moreover, multiplications only involve layers inputs and weights.
Because the weights are quantized to take as values powers of two, all the multiplications required by the \gls{NN} correspond to bit shifts.
As the weights are assumed to be fixed after deployment, the bit shifts can be ``hardwired'' in the hardware implementation, removing the need for storing the weights in memory, as well as programmable bit shifters.

The \gls{ML}-receiver is assumed to be implemented by measuring the squared Euclidean distance of the received signal with each of the $M$ possibly sent signals, and taking the closest.
Therefore, it requires squaring operations, i.e., multiplications.
On a fixed-point system, each multiplication requires $K-1$ additions as well as $K$ bit shifts, these latter being assumed to have a negligible complexity compared to additions.
Accordingly, only additions are considered to compare the complexities of the implementations.
Complexities of the considered schemes are therefore evaluated by comparing the number of required additions.
Table~\ref{tab:complex} shows the number of additions required by the quantized \gls{NN}-based receiver and the \gls{ML} receiver, for which each multiplication was counted as $K-1$ additions.
Notice that the complexity of an addition depends on how it is implemented, and of the number of bits $K$ used in the fixed-point system.
As one can see, the quantized \gls{NN}-based receiver requires approximately $60\%$ less additions than the \gls{ML} receiver with $K = 14\:$bits, without incurring significant \gls{BLER} degradation as seen in the previous section.
This encouraging result illustrates how \gls{NN}-based approaches have the potential to significantly reduce the complexity of communication systems, without significant loss of performance.

\begin{table}[ht]
\caption{Number of additions required by the quantized \gls{NN}-based receiver and \gls{ML} receiver \label{tab:complex}}
\begin{tabular}{c|ccc}
                        & For any $K$            & $K=14$   & Complexity -- $K=14$    \\
\hline
\gls{ML} receiver       & \makecell{$2048(K-1)$\\$ + 3840$}       & $30464$      & $100\%$            \\
\hline
\gls{NN}-based receiver & $10496$                & $10496$      & $34.5\%$            \\
\end{tabular}
\end{table}

\section{Conclusion}
\balance
We presented in this paper an approach to reduce the implementation complexity of \gls{NN}-based communication algorithms.
Considering an \gls{NN}-based receiver as example, complexity reduction was achieved by quantizing the weights so that they take as values powers of two, reducing all multiplication to bit shifts in fixed-point arithmetic.
Compared to naive direct compression, this approach incurs almost no \gls{BLER} increase, while enabling significant gain in computational complexity.
Our results show that the compressed \gls{NN}-based receiver achieves \glspl{BLER} close to the ones of \gls{ML} detection, while enabling $60\:\%$ gains in computational complexity when implemented on a 14~bits fixed-point arithmetic system.

We believe that future work on quantization, compression, and more broadly, the efficient hardware implementation of \glspl{NN} for physical layer tasks is required by our community before machine learning-based solutions can make it into commercial products.

\section*{Acknowledgment}
The authors thank Luc Dartois for comments that greatly improved the manuscript.

\bibliographystyle{IEEEtran}
\bibliography{IEEEabrv,bibliography}
\pagebreak
\end{document}